\documentclass[10pt,conference]{IEEEtran}
\IEEEoverridecommandlockouts
\usepackage{cite}
\usepackage{amsmath,amssymb,amsfonts}
\usepackage{algorithm}
\usepackage{algpseudocode}
\usepackage{graphicx}
\usepackage{textcomp}
\usepackage{xcolor}
\def\BibTeX{{\rm B\kern-.05em{\sc i\kern-.025em b}\kern-.08em
    T\kern-.1667em\lower.7ex\hbox{E}\kern-.125emX}}

\usepackage{stfloats}
\usepackage[colorlinks,bookmarksopen,bookmarksnumbered,citecolor=red, urlcolor=black]{hyperref}
\usepackage[capitalize]{cleveref}
\hypersetup{colorlinks=true,breaklinks=true}

\usepackage{tcolorbox}

\newboolean{showcomments}
\setboolean{showcomments}{true}
\ifthenelse{\boolean{showcomments}}
{\newcommand{\nbc}[3]{
 {\colorbox{#3}{\bfseries\sffamily\scriptsize\textcolor{white}{#1}}}
 {\textcolor{#3}{\sf\small$\blacktriangleright$\textit{#2}$\blacktriangleleft$}}
 }
}
{\newcommand{\nbc}[3]{}
 }

\begin{document}

\title{Context-Aware CodeLLM Eviction for AI-assisted Coding}

\author{
    \IEEEauthorblockN{Kishanthan Thangarajah\IEEEauthorrefmark{1}, Boyuan Chen\IEEEauthorrefmark{1}, Shi Chang\IEEEauthorrefmark{2},
    Ahmed E. Hassan\IEEEauthorrefmark{3}}
    \IEEEauthorblockA{\IEEEauthorrefmark{1}Centre for Software Excellence, Huawei Canada}    \IEEEauthorblockA{\IEEEauthorrefmark{2}Western University, Canada, \IEEEauthorrefmark{3}Queen's University, Canada} 
    \IEEEauthorblockA{\texttt{cse@huawei.com}}
}

\maketitle

\begin{abstract}
AI-assisted coding tools powered by Code Large Language Models (CodeLLMs) are increasingly integrated into modern software development workflows. To address concerns around privacy, latency, and model customization, many enterprises opt to self-host these models. However, the diversity and growing number of CodeLLMs, coupled with limited accelerator memory, introduce practical challenges in model management and serving efficiency. This paper presents \textsc{CACE}, a novel context-aware model eviction strategy designed specifically to optimize self-hosted CodeLLM serving under resource constraints. Unlike traditional eviction strategies based solely on recency (e.g., Least Recently Used), \textsc{CACE} leverages multiple context-aware factors, including model load time, task-specific latency sensitivity, expected output length, and recent usage and future demand tracked through a sliding window. We evaluate \textsc{CACE} using realistic workloads that include both latency-sensitive code completion and throughput-intensive code reasoning tasks. Our experiments show that \textsc{CACE} reduces Time-to-First-Token (TTFT) and end-to-end (E2E) latency, while significantly lowering the number of model evictions compared to state-of-the-art systems. Ablation studies further demonstrate the importance of multi-factor eviction in balancing responsiveness and resource efficiency. This work contributes practical strategies for deploying scalable, low-latency AI coding assistants in real-world software engineering environments.
\end{abstract}


\section{Introduction}
\label{sec:intro}
CodeLLMs are a distinct category of Large Language Models (LLMs) specifically designed for coding tasks. They are trained on a vast number of repositories of code artifacts and related documentation \cite{jiang2024survey, xu2022systematic}. AI-assisted coding with CodeLLMs has significantly reshaped software development, boosting productivity by reducing both coding time and errors~\cite{ziegler2024measuring, cursor_editor, copilot_editor, murali2024ai, aws_q_dev_editor, intellij_editor}. AI-assisted coding enables developers to concentrate on higher-level design and problem-solving tasks rather than boilerplate implementation. With the rapid release of new models, the number and variety of available CodeLLMs continue to grow. However, not all models perform equally across different coding scenarios. Some CodeLLMs excel at generating code, while others are more effective at understanding or reasoning through complex logic, especially when tailored to specific programming languages~\cite{zheng2023codegeex, gu2025domain, xu2024cruxeval, chai2024mceval, Mellum-4b-base, jetbrains_full_line}. Likewise, performance can vary across application domains such as web development, mobile, blockchain, game, distributed systems, or networking, where certain CodeLLMs demonstrate domain-specific strengths~\cite{gu2025effectiveness, zhu2025domaineval, li2024evocodebench, zheng2024well}. As a result, developers now have the flexibility to choose from a wide range of models optimized for their particular use cases, making model selection increasingly critical to achieving high performance and relevance in AI-assisted coding workflows.

To boost developer productivity, many organizations are integrating AI-assisted coding tools powered by CodeLLMs~\cite{semenkin2024full, murali2024ai, dunay2024multi, cursor_editor}. Enterprise software across diverse domains often rely on developers working with multiple programming languages~\cite{mayer2015empirical, li2021multi, kochhar2016large, yang2024multi, cassano2022multipl, dev_blogpost_multipl}. For instance, a typical full-stack application may involve front-end developers using JavaScript or TypeScript, back-end developers writing services in Java or Python, DevOps engineers authoring infrastructure code in Go, and automation tasks scripted in Bash. This diversity is driven by the specialization of tools and languages for different layers of the software stack. Consequently, an effective CodeLLM-based assistant must support multiple languages and provide fast, high-quality responses across a broad spectrum of coding tasks and codebases.

In addition, due to growing privacy and security concerns~\cite{eu_ai_act, wu2024unveiling, das2025security}, these organizations are increasingly opting to self-host CodeLLMs, including fine-tuned ones on proprietary data, rather than relying on external model APIs. While self-hosting specialized CodeLLMs enhances privacy control and allows customization to meet enterprise-specific requirements, it introduces a new challenge: managing limited in-house accelerator resources in environments where the total number of CodeLLMs needed in use far exceeds available resources. As developers expect near-instantaneous feedback, especially for latency-critical tasks such as code completion, reducing the cold-start latency of LLM requests becomes essential. One common approach for existing LLM serving systems is to enable model multiplexing. 

Model multiplexing refers to concurrently hosting and dynamically managing multiple models on shared accelerator resources. This approach optimizes resource usage by loading, unloading, and routing inference requests among various model instances according to real-time demands. Frameworks such as Ray Serve~\cite{ray_multiplex}, Ollama~\cite{ollama}, ModelMesh~\cite{model_mesh}, and other state-of-the-art systems~\cite{yu2024faaswap} exemplify this approach by dynamically loading, unloading, and routing inference requests among multiple model instances within constrained accelerator memory. These systems aim to maximize resource utilization and throughput, typically employing simple heuristics, such as Least Recently Used (LRU), for model eviction decisions. However, without context-aware strategies, they often encounter unnecessary data movement and cold-start latencies that negatively impact developer experience in latency-sensitive tasks. This makes efficient model eviction strategies a key requirement to ensure the most relevant models are loaded dynamically, cold starts are minimized, and both latency and resource costs are optimized.

In this paper, we introduce \underline{\textbf{C}}ontext-\underline{\textbf{A}}ware \underline{\textbf{C}}odeLLM \underline{\textbf{E}}viction (\textsc{CACE}), a dynamic and context-aware model eviction policy designed specifically for efficient and responsive CodeLLM serving in enterprise environments. Unlike traditional eviction methods, such as LRU, \textsc{CACE} integrates multiple context-aware factors, including model loading times, recent usage patterns, task-specific latency requirements, and predicted future developer requests. Through a comprehensive evaluation using realistic developer workloads comprising both latency-sensitive code completion tasks and throughput-intensive code reasoning tasks, we demonstrate that \textsc{CACE} significantly reduces Time-To-First-Token (TTFT) and end-to-end (E2E) latency, while minimizing unnecessary data movement (i.e., model reload). Additionally, we conduct extensive ablation studies to isolate and analyze the relative impact of each individual factor, providing clear insights into the effectiveness of our multi-factor eviction strategy.

Specifically, we explore the following research questions (RQs):
\begin{itemize}
    \item \textbf{RQ1:} Can \textsc{CACE} reduce the overall number of model evictions compared to conventional eviction strategies in a self-hosted CodeLLM environment?
    \item \textbf{RQ2:} How effectively does \textsc{CACE} reduce TTFT and E2E latency in a self-hosted CodeLLM environment?
    \item \textbf{RQ3:} What is the individual impact of each context-aware factors within the \textsc{CACE} eviction policy?
\end{itemize}

To ensure transparency and reproducibility, we intend to make our solution and all associated datasets publicly available~\cite{cace_replication}. The open-source release will allow reviewers and researchers to evaluate, replicate, and build upon our work.

The remainder of this paper is structured as follows: Section~\ref{sec:background} presents the background and motivation for efficient, latency-sensitive CodeLLM serving in enterprise environments. Section~\ref{sec:challenges} highlights key operational challenges and industry trends. Section~\ref{sec:solution} describes the detailed design of our proposed \textsc{CACE} approach, including the eviction policy and algorithmic factors. Section~\ref{sec:research_questions} outlines the experimental methodology, datasets, workload patterns, and evaluation metrics. Section ~\ref{sec:discussion} discusses about insights from the results potential future research directions. Section~\ref{sec:related_work} focus on related work, positioning our contributions within the landscape of existing solutions, including model multiplexing and memory management strategies. Section~\ref{sec:threat_to_validity} explores potential threats to validity and limitations of our approach. Finally, Section~\ref{sec:conclusion} concludes the paper and outlines promising directions for future research.

\section{Background and Motivation}
\label{sec:background}
\subsection{Latency in AI-Assisted Coding}
AI-assisted coding powered by CodeLLMs significantly enhances developer productivity through intelligent suggestions for tasks such as code completion, test generation, and codebase understanding. In particular, within interactive development environments (IDEs), developers have stringent expectations regarding responsiveness and latency. Even minor delays in tasks such as code completion can disrupt workflow continuity, making latency a critical barrier to the effective adoption of coding assistants~\cite{dunay2024multi}. Systems such as CodeCompose at Meta~\cite{murali2024ai} address this challenge by using a fine-tuned version of CodeLlama, delivering multi-line suggestions within milliseconds, aligning closely with developers' fast-paced workflow. Similarly, DeepVulGuard \cite{steenhoek2024closing} employs a small finetuned CodeBERT model to rapidly identify vulnerabilities. The Cursor team has used a Mixture-of-Expert based model to handle large codebases and long input context length while reducing inference latency~\cite{cursor_podcast} and lastly, JetBrains IDEs integrate lightweight, language-specific models optimized for low-latency, line-level code completions~\cite{semenkin2024full}. These industry efforts collectively underscore the importance of minimizing inference latency to meet developer expectations for real-time assistance.

\subsection{Motivations for Self-Hosting CodeLLMs}
Enterprises deploying AI-assisted coding tools face additional considerations beyond latency, particularly about data privacy~\cite{wu2024unveiling, li2023multi}, security ~\cite{rathod2025privacy, das2025security}, and operational control. Due to stringent privacy regulations such as GDPR~\cite{eu_ai_privacy_protection} and the necessity of safeguarding proprietary source code~\cite{eu_ai_act}, many organizations prefer self-hosted CodeLLMs rather than relying on external API services. Self-hosting allows enterprises to control data exposure risks and reduces latency by eliminating network calls to external servers, thus ensuring more responsive and secure interactions for developer tasks.

\subsection{Selecting the Right CodeLLM for the Task}
An inherent complexity arises from the diverse range of coding tasks and the availability of numerous CodeLLM variants optimized for specific use cases. Small, specialized CodeLLMs, fine-tuned per programming language (e.g., Python, Java, C++), typically support latency-critical tasks like code completion. Conversely, larger models are utilized for computationally intensive reasoning tasks, which, although less latency-sensitive, require greater computational resources. Model effectiveness varies significantly across languages and domains. For example, Qwen2.5-Coder-32B excels in Python and C\#, yet underperforms in Java, C++ and PHP~\cite{hui2024qwen2}, while Codestral 20B surpasses larger models in specific languages such as SQL and Swift~\cite{codestral2024}. For code reasoning tasks, such as code output reasoning, WizardCoder 15B performs well for Java and Scala, Starcoder2 15B performs well for Swift, Codegen 6B for Shell, and CodeLlama 34B performs well for Rust, C++, and Julia languages~\cite{xu2024cruxeval}. Such variability necessitates providing developers with a choice of models tailored explicitly to their tasks and development contexts.

\subsection{Necessity of Dynamic CodeLLM Eviction}
Accommodating all these models simultaneously on limited accelerator resources (GPUs/TPUs/NPUs) is impractical due to both cost and accelerator memory constraints. Traditional model serving approaches with model multiplexing, often rely on simple eviction policies, such as Least Recently Used (LRU), which do not adequately account for the unique characteristics of coding tasks, including latency sensitivity, model load time, and anticipated developer usage patterns. Consequently, naive eviction strategies can frequently trigger unnecessary model reloads, introducing disruptive cold-start latencies and diminishing the efficiency and responsiveness of the development workflow.

\subsection{Existing Tools and Infrastructure}
Recognizing these practical challenges, recent industry and academic work has introduced frameworks like Ray Serve~\cite{ray_multiplex}, Ollama~\cite{ollama}, and FaaSwap~\cite{yu2024faaswap}, each with distinct capabilities in concurrency support, model multiplexing, and dynamic loading mechanisms. However, these frameworks typically do not integrate comprehensive task-aware eviction logic, often defaulting to generic heuristics that neglect critical context factors, leading to suboptimal resource management and responsiveness.

Addressing these gaps requires a novel, context-aware approach capable of dynamically prioritizing model evictions by considering multiple factors: recent usage patterns, task-specific latency requirements, model load times, and anticipated developer requests. 

\section{Challenges}
\label{sec:challenges}
Current production-grade LLM serving systems, such as Ray Serve~\cite{ray_multiplex}, Ollama~\cite{ollama}, ModelMesh~\cite{model_mesh}, and state-of-the-art research frameworks like FaaSwap~\cite{yu2024faaswap} primarily rely on simplistic LRU policies for model eviction between host and accelerator memory. While LRU is straightforward and easy to implement, it often triggers unnecessary model evictions, as it fails to deal with the following challenges. As a result, it leads to increased data movement (i.e., the excessive model weights) and elevated latency, particularly in TTFT, which negatively impacts developer experience.

\subsection{Minimizing Data Movement Overheads}

Accelerators such as GPUs, TPUs, and NPUs are expensive to provision and maintain, resulting in limited availability within enterprises. Simultaneously, the number and diversity of CodeLLM variants, including multiple fine-tuned models tailored for different tasks and programming languages continues to grow. In modern enterprise software development, developers across various roles, such as front-end engineers, back-end engineers, and DevOps teams often require access to different models concurrently to support their workflows. However, the total accelerator memory is typically insufficient to accommodate all required models at once, necessitating frequent model evictions.

Consequently, model multiplexing between host and accelerator memory becomes essential. Traditional eviction strategies, such as LRU, rely solely on access recency and overlook key model characteristics. Importantly, CodeLLMs vary widely in size and load time, and indiscriminate eviction can lead to substantial cold-start delays. For example, the cost of evicting a large model like DeepSeek-R1 is different from a specialized fine-tuned Qwen-Coder-3B. An effective eviction policy must therefore consider model size, load time, and predicted usage frequency to minimize data movement and ensure timely model availability.

\subsection{Managing Task-Specific Latency Requirements}

Different types of AI-assisted coding tasks impose distinct latency requirements that directly impact the developer experience. Code completion tasks are highly sensitive to TTFT latency, as developers expect rapid, near-instantaneous suggestions while typing. These tasks typically involve short output lengths and demand sub-second responsiveness to maintain an uninterrupted development flow. In contrast, code reasoning tasks, such as multi-line code synthesis or complex bug explanation, can tolerate higher E2E latency due to the inherently longer outputs and deeper computation involved. However, even for E2E-sensitive tasks, response time must remain within an acceptable threshold to avoid disrupting the developer's focus or productivity.

With multiple CodeLLMs competing for constrained accelerator memory, it becomes crucial to design model retention and eviction policies. That is to say, when accelerator memory is full and there is an incoming model request, we have to strategically select an existing loaded model to evict.
For example, naively evicting a model serving a TTFT-critical task in favor of a less urgent one can significantly degrade the real-time responsiveness required for code completion. Conversely, excessive eviction of large reasoning models may inflate E2E latency due to high reload overhead. 

\section{Context-Aware CodeLLM Eviction (\textsc{CACE})}
\label{sec:solution}
\begin{figure*}[t]
    \centering
    \includegraphics[width=0.9\linewidth]{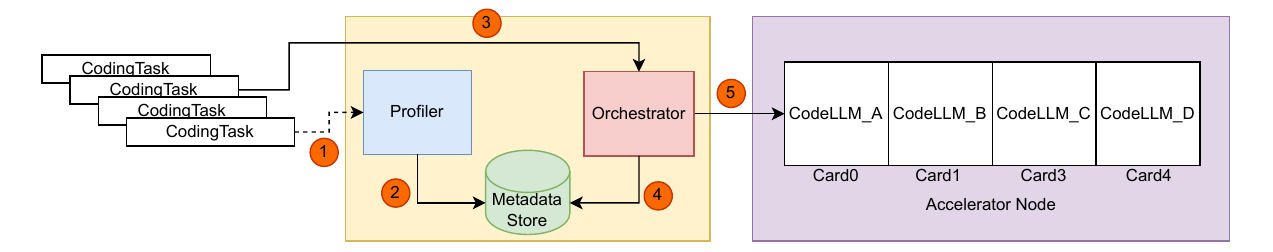}
    \caption{The Overall Workflow of \textsc{CACE}}
    \label{fig:CACE}
\end{figure*}

To enhance overall model eviction performance in serving systems, \textsc{CACE} is designed to minimize the number of model evictions, thereby reducing TTFT and E2E latencies. We first describe the system design in Section~\ref{subsec:system_arch}, and then further break down the details on the core factors that impact the eviction strategy in Section~\ref{subsec:eviction_strategy}.

\subsection{\textsc{CACE} System Architecture and Serving Pipeline}
\label{subsec:system_arch}

The \textsc{CACE} solution is integrated into a single-node CodeLLM serving architecture designed to meet the demands of enterprise software development environments. These environments require both low-latency responses for interactive tasks like code completion and support for long-running reasoning tasks. \textsc{CACE} serves as an intelligent, context-aware model eviction technique to manage accelerator memory efficiently under dynamic workloads.

The serving pipeline, as described in Figure~\ref{fig:CACE}, begins with a model registration step \textcircled{1}, where developers register a new instance of CodeLLMs into the system. Upon registration, the \textbf{Profiler} loads the model in a staging area and records static metrics such as model load time and memory usage. These metrics are stored in the~\textbf{Metadata Store}, which is backed by a persistent database. This is an offline step before the model is made available for serving \textcircled{2}.

At runtime (online), the coding tasks requests are received \textcircled{3} by the ~\textbf{Orchestrator}. This will query the ~\textbf{Metadata Store} to retrieve relevant profiling data for models residing in the accelerator memory, which can be evicted \textcircled{4}. It then combines this information with real-time request context, such as task latency sensitivity (e.g., TTFT vs. E2E), expected output token length, and recent model usage, to compute an eviction score for each model currently residing in the accelerator memory. This score is used to guide model eviction decisions(described in Section~\ref{subsec:eviction_strategy}), ensuring that latency-critical and frequently used models are prioritized, while less critical models are evicted to make room for incoming ones. The ~\textbf{Orchestrator} make sure that the requested CodeLLM is loaded in the accelerator and ready before the request is dispatched \textcircled{5}. By decoupling profiling from runtime execution and leveraging persistent metadata access, \textsc{CACE} enables task-sensitive, context-aware model management optimized for resource-constrained accelerator environments.

\begin{algorithm} [H]
\caption{\textsc{CACE} Algorithm}
\label{alg:CACESwap}
\begin{algorithmic}[1]
\State $L1 \gets$ Models in Accelerator memory
\State $L2 \gets$ Models to be loaded
\State $w \gets$ Sliding window length
\Procedure{\textsc{CACE}}{$L1, L2, len$}
    \State Remove duplicates from $L2$
    \State Sort $L1$ by last accessed time
    \State $S \gets \{\}$ \Comment{Eviction scores}
    \ForAll{$m \in L1$}
        \State $o \gets$ \Call{GetExpectedOutputTokens}{$m$}
        \State $S[m] \gets$ \Call{CalcEvictionScore}{$m, L2, w, o$}
    \EndFor
    \State $m_{evict} \gets \arg\max S[m]$
    \State \Call{Evict}{$m_{evict}$}
\EndProcedure
\end{algorithmic}
\end{algorithm}

Unlike traditional approaches that rely solely on LRU policy, our method integrates multiple sources of information to guide model eviction decisions. We will break down our method next.

\subsection{Eviction Score Calculation}
\label{subsec:eviction_strategy}

To guide evictions in our evicting algorithm, we introduce an \textit{Eviction Score} that combines multiple factors influencing model retention. The score is designed with a few considerations.
First, it incorporates anticipated future load requests by monitoring the incoming request queue from the serving system, enabling prioritization of models likely to be required imminently. This information is captured as a list L2 (models to be loaded). Second, this knowledge pertains to the set of models already loaded into accelerator memory. This information is captured as a list L1 (models already loaded in accelerator memory). We maintain a sliding window containing a fixed number of user-requested and currently loaded models within this window. As new requests arrive, the window dynamically slides to update the set of relevant models. Third, the algorithm accounts for model load times, assigning higher weights to larger models whose reloading incurs greater latency penalties. Finally, unlike previous methods, our approach integrates task criticality by evaluating the expected output token count associated with each request, thereby prioritizing models critical for latency-sensitive tasks such as code completion. By synthesizing these factors, the proposed policy aims to optimize evicting decisions for improved responsiveness and resource utilization.

Let:
\begin{itemize}
    \item $t$: last used time of the model in L1 (timestamp),
    \item $l$: model load time (in seconds),
    \item $w$: sliding window length,
    \item $i$: index position of the model in L2,
    \item $o$: expected number of output tokens,
    \item $w_1$: weighting factor for output tokens.
\end{itemize}

\[
\text{EvictionScore} \;=\;
\frac{1}{1+\ln t}
\;+\;
\frac{1}{1+\tfrac{l}{100}}
\;+\;
\frac{i}{w} 
\;+\;
w_1 \cdot o
\]

Based on these design choices, the eviction score incorporates four key factors, \textbf{P1 (Recency)}, \textbf{P2 (Reloading Cost)}, \textbf{P3 (Future Demand)}, and \textbf{P4 (Task Criticality)}, each contributing to a more informed and context-sensitive eviction policy, the rationale behind the formula are explained as follows:

\begin{itemize}
    \item \textbf{P1: Recency (Last Used Time in L1)} \\
    This factor captures how recently a model was accessed in the accelerator memory. A model used more recently will have a smaller $t$ (last used time of the model in L1) value, yielding a larger inverse score contribution. This protects frequently accessed models from eviction. In contrast, models with older access times contribute less to the score, increasing their likelihood of being evicted.

    \item \textbf{P2: Reloading Cost (Model Load Time)} \\
    This factor reflects the time required to reload a model into the accelerator memory, using $l$ (model load time) as a proxy. Models with shorter load times (typically smaller models) yield higher contributions to the score, indicating they are cheaper to evict. Models with longer load times are penalized less, making their eviction less favorable due to higher reloading overhead.

    \item \textbf{P3: Future Demand (Sliding Window Forecast)} \\
    This factor estimates the urgency of future model usage based on the position of the model in the future request window. It is calculated as the fraction between model’s $i$ (index position of the model in L2) and the $w$ (which is the sliding window length). A smaller value for this factor implies the model is going to be reuses soon, resulting in increased retention priority. Models needed further in the future contribute less to the score and are more likely to be evicted. If the model is not present in the L2, then a default value of $1$ is given.

    \item \textbf{P4: Task Criticality (Expected Output Tokens)} \\
    This factor reflects task sensitivity by using the expected number of output tokens, weighted by a tunable parameter $w_1$. Larger token counts that are common in E2E-centric tasks, like code reasoning, contribute more to the score. Lower values indicate latency-critical tasks such as code completion, thereby prioritizing models that serve those tasks for retention.
\end{itemize}

By integrating these four factors, the eviction policy of \textsc{CACE} goes beyond simple recency-based strategies to make context-aware decisions that reflect both system constraints and workload context. This aims to provide more efficient model retention and improved performance under constrained accelerator resources.

\section{Evaluation}
\label{sec:research_questions}
In this section, we first describe the case study setup in Section~\ref{subsec:case_study_setup}, then we answer the three RQs respectively (\crefrange{subsec:rq1}{subsec:rq3}).

\subsection{Case Study Setup}
\label{subsec:case_study_setup}
All experiments are conducted on an Ubuntu server equipped with an AMD Ryzen Threadripper PRO 5975WX (64 cores) and four accelerators with 48 GB of memory. We used vLLM as the inference engine and dedicated a single card for each model instance. At the same time, only four model instances will be loaded simultaneously with four accelerators. We used Ray Serve~\cite{ray_multiplex} as the baseline system, which implements an LRU-based model multiplexing strategy, similar to RayServe and FaaSwap.
We then evaluate \textsc{CACE} against LRU with two benchmark datasets:

\begin{itemize}
    \item \textbf{McEval:} A benchmark for multilingual code completion tasks~\cite{chai2024mceval}.
    \item \textbf{CRUXEVAL-X:} A benchmark for multilingual code reasoning tasks ~\cite{xu2024cruxeval}.
\end{itemize}

For code completion tasks that are highly sensitive to TTFT, we assign a small CodeLLM with 500M parameters per programming language, following the approach of Semenkin et al.~\cite{semenkin2024full}, who adopted fine-tuned and lightweight models for individual languages at JetBrains. For code reasoning tasks, which are more sensitive to E2E latency, we utilized large CodeLLMs with 7B parameters for each programming language to better support complex analysis and multi-step inference. We selected eight programming languages based on their consistent ranking in industry-recognized sources such as the RedMonk and TIOBE indexes~\cite{redmonk2024, tiobe2025}: Java, Python, C++, C, Go, Rust, C\#, and JavaScript. Accordingly, our experimental setup includes eight small CodeLLMs and eight large CodeLLMs, one of each per language.

To evaluate the effectiveness of our approach, we define three workload request arrival patterns that reflect realistic usage scenarios~\cite{code_sginal_blog, jetbrains_blogpost, so_blogpost} in enterprise-grade software development environments:

\begin{itemize}
    \item \textbf{Pattern 1: Uniform.} This pattern simulates an equal distribution of code completion and code reasoning requests (50\% each) across eight programming languages. It serves as a balanced baseline for evaluating the \textsc{CACE} eviction policy under non-skewed conditions.

    \item \textbf{Pattern 2: IDE Heavy.} An imbalanced workload with 70\% code completion and 30\% code reasoning requests. This is to simulate a real-world use case where developers often call code completion in the IDE's compared to code reasoning tasks.

    \item \textbf{Pattern 3: Language Popularity Skewed.} A variation of Pattern 2, and reflects real-world language usage trends. Within the 70\%-30\% allocation for each coding task type, 20\% of requests are allocated to Java, 20\% to Python, 20\% to C++, 20\% JavaScript, and the remaining 20\% are distributed among Go, Rust, C, and C\#. This distribution is informed by recent programming language popularity data, where Python, Java, JavaScript, and C++ consistently rank among the top-most-used languages. Specifically, the RedMonk ~\cite{redmonk2024} and TIOBE ~\cite{tiobe2025} lists Python, C++, and JavaScript as the top-ranked languages, with Go, Rust, and JavaScript also featuring within the top 20.
\end{itemize}

These patterns enable us to assess the adaptability and efficiency of \textsc{CACE} across both balanced and skewed usage scenarios, ensuring its applicability to a range of real-world development workloads.

Lastly, for every workload pattern, we synthetically generate request streams using a Poisson arrival distribution over a fixed $30$ second time window. Each request is then labeled according to the target pattern (Uniform, IDE Heavy, or Popularity Skewed). Before dispatching the stream to the serving stack, we shuffle all requests to avoid bursty language or task-specific clusters and to ensure a realistic, interleaved mix of code-completion and code-reasoning tasks within the 30-second interval.

\subsubsection*{Notation for \textsc{CACE} Variants}
To systematically evaluate the impact of each factor in our eviction policy, we define the following notation to distinguish between different variants of \textsc{CACE}:

\begin{itemize}
    \item \textbf{CACE$^{\ast\ast}$}: The full eviction strategy with all four factors enabled ((P1) Recency, (P2) Reloading Cost, (P3) Future Demand, and (P4) Task Criticality).
    \item \textbf{CACE$^{-P1}$}: Variant with \textbf{Recency} disabled.
    \item \textbf{CACE$^{-P2}$}: Variant with \textbf{Reloading Cost} disabled.
    \item \textbf{CACE$^{-P3}$}: Variant with \textbf{Future Demand} disabled.
    \item \textbf{CACE$^{-P4}$}: Variant with \textbf{Task Criticality} disabled.
\end{itemize}

These variants are used across RQs to analyze the contribution of each factor to eviction effectiveness, latency reduction, and cache performance.

\subsection{RQ1: Can \textsc{CACE} reduce the overall number of model evictions compared to conventional strategies in a self-hosted CodeLLM environment?}
\label{subsec:rq1}
\subsubsection{Hypothesis}

We hypothesize that \textsc{CACE} will significantly reduce the total number of model evictions compared to traditional LRU-based eviction strategies by avoiding unnecessary evictions and retaining frequently used models in accelerator memory. While this research question focuses on eviction behavior, a reduction in model evictions may also contribute to improved latency characteristics by lowering the frequency of cold-start loading events, which will be discussed in the next RQ.

\subsubsection{Results}

For this RQ, we use the setup as described in Section~\ref{subsec:case_study_setup}.
In addition, we disable the Task Criticality weighting factor in the eviction score equation to isolate and evaluate the performance of model eviction alone.

We evaluate system performance based on the following metrics:

\begin{itemize}
    \item \textbf{Model Cache Hit Rate:} Among the total number of requests being served, the proportion of requests served from models already loaded in accelerators.
    \item \textbf{Model Load Overhead:} Measure of time spent on loading models (highlighting the unproductive time spent on loading).
\end{itemize}

\begin{figure}[ht]
    \centering
    \includegraphics[width=\linewidth]{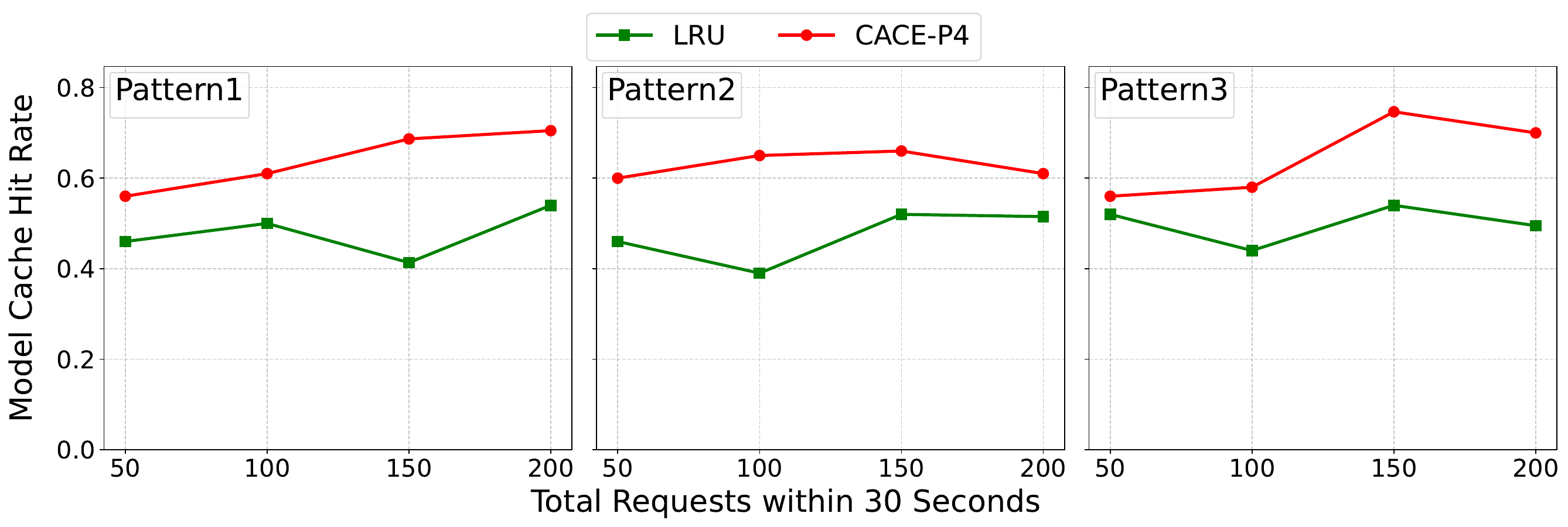}
    \caption{Comparing Model Cache Hit Rate against different request load patterns of CACE$^{-P4}$ and LRU for RQ1.}
    \label{fig:rq1_cache_hit_rate}
\end{figure}

\begin{figure}[ht]
    \centering
    \includegraphics[width=\linewidth]{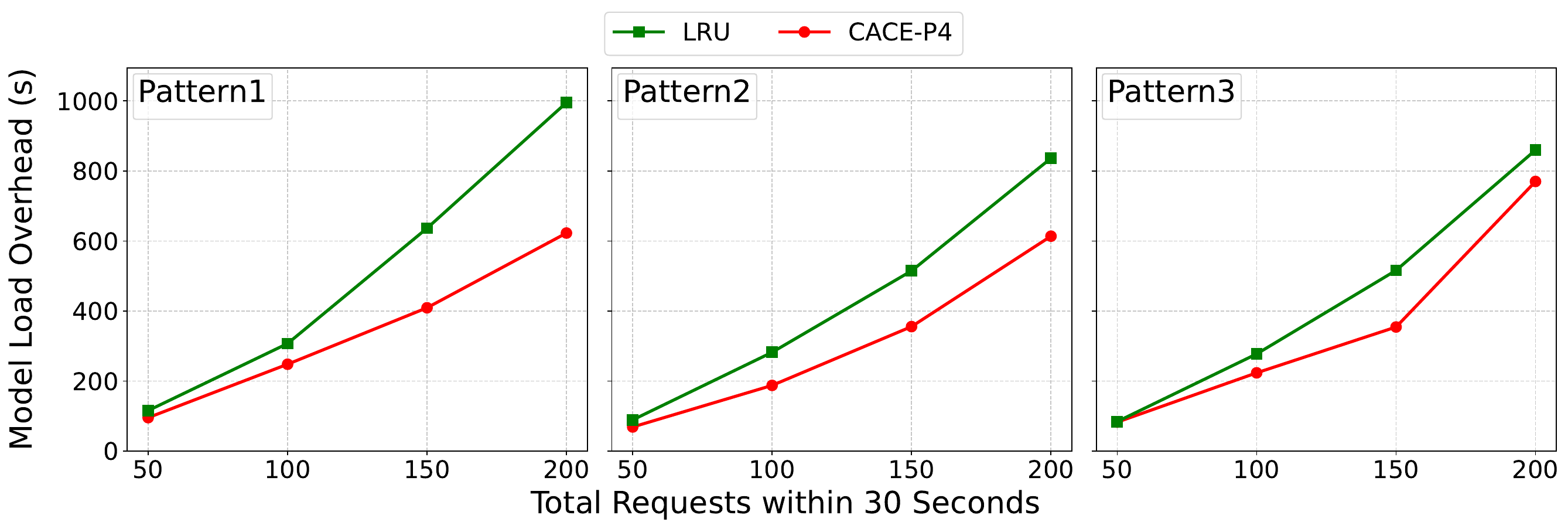}
    \caption{Comparing Model Load Overhead against different request load patterns of CACE$^{-P4}$ and LRU for RQ1.}
    \label{fig:rq1_model_load_overhead}
\end{figure}

Figures~\ref{fig:rq1_cache_hit_rate} and~\ref{fig:rq1_model_load_overhead} present the comparison between CACE$^{-P4}$ (with Task Criticality disabled) and the baseline LRU strategy under three distinct workload patterns. Figure~\ref{fig:rq1_cache_hit_rate} demonstrates that CACE$^{-P4}$ consistently achieves higher Model Cache Hit Rate across all workloads compared to LRU. 

Similarly, Figure~\ref{fig:rq1_model_load_overhead} reveals that CACE$^{-P4}$ significantly reduces cumulative model load overhead compared to LRU. For instance, under Pattern 2, CACE$^{-P4}$ reduces the total Model Load Overhead by about \textbf{27\%}, and under Pattern 3, by about \textbf{31\%}. 

\begin{tcolorbox}[colback=white, colframe=black, boxrule=1pt, arc=0pt, left=10pt, right=10pt, top=10pt, bottom=10pt]
\textbf{Findings:}
Even without the Task Criticality factor, CACE$^{-P4}$ reduces Model Load Overhead by up to 31\% and increases Model Cache Hit Rate by up to 41\% compared with a pure LRU policy. These improvements shows the effectiveness only by considering Recency, Reloading Cos, and Future Demand in eviction decisions to reduce unnecessary data movements.

\textbf{Implication:}
A recency-only strategy is insufficient in multi-model CodeLLM serving. Designers should incorporate at least model reload cost and short-term demand forecasting, and the choice of additional factors, such as latency, can then be aligned with performance goals. We further explore this in the next research question.
\end{tcolorbox}

\subsection{RQ2: How effectively does \textsc{CACE} reduce TTFT and E2E
latency in a self-hosted CodeLLM environment?}
\label{subsec:rq2}
AI-assisted coding tasks such as inline code completions are highly sensitive to TTFT latency, as developers expect near-instantaneous responses. In addition, complex tasks such as large scale code refactoring can tolerate long TTFT but still benefit from reduced E2E latency. Traditional model eviction strategies like LRU are agnostic to task specific latency requirements, which is unique in CodeLLM generations. For instance, LRU might fail to accommodate TTFT when deciding to evict a model just because it is not the most recently used model.

In this RQ, we evaluate \textsc{CACE}'s eviction policy, which incorporates the Task Criticality factor, such as expected output length and latency sensitivity, by dynamically monitoring the request queue. This approach enables \textsc{CACE} to prioritize retaining models that are critical to latency-sensitive tasks, thereby reducing the frequency of cold starts and improving both TTFT and E2E latency metrics.

\subsubsection{Hypothesis}

We hypothesize that \textsc{CACE} will significantly reduce TTFT for latency-sensitive tasks such as code completion, and lower E2E latency for tasks like code reasoning, at the same time not compromising on the model eviction performance, when compared to conventional LRU-based eviction strategies. This improvement is expected to manifest in:
\begin{itemize}
    \item Reduced TTFT for short-response tasks.
    \item Reduced E2E latency for complex, long-response tasks.
    \item Fewer model evictions. Expect this to be lower than that in RQ1. 
\end{itemize}

\subsubsection{Results}

We use the same experiment setup as described in Section~\ref{subsec:case_study_setup} with the same three request arrival patterns under realistic usage scenarios to test latency sensitivity. We then evaluate system performance based on the following metrics:

\begin{itemize}
    \item \textbf{TTFT:} Time-to-first-token for code completion tasks.
    \item \textbf{E2E Latency:} Total end-to-end latency for code reasoning tasks.
    \item \textbf{Model Cache Hit Rate:} Percentage of requests served without requiring a model reload.
\end{itemize}

TTFT and E2E latency are evaluated to showcase the benefits of our approach, while Model Cache Hit Rate is used to explain the rationale behind the improvements.

\begin{figure}[ht]
    \centering
    \includegraphics[width=\linewidth]{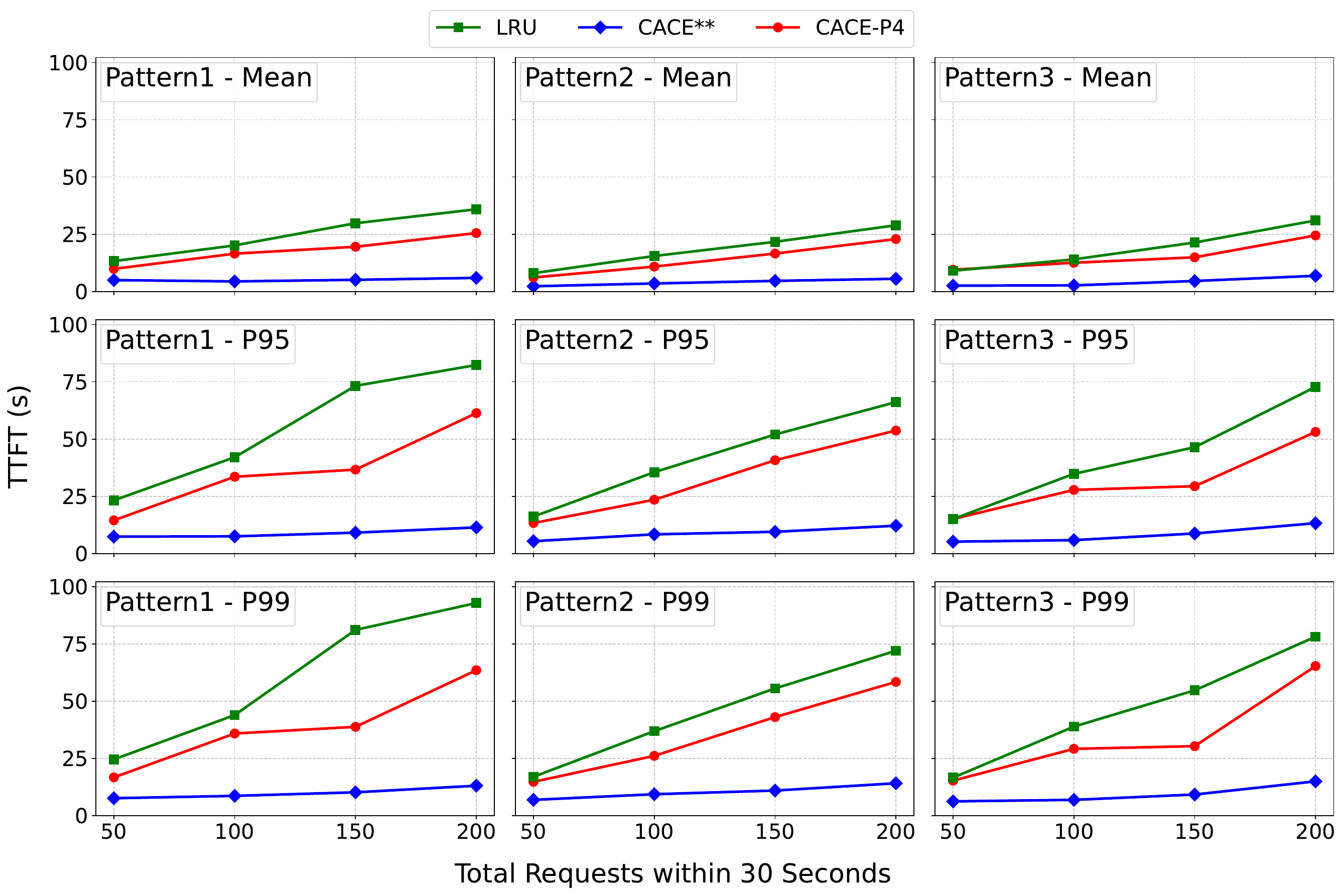}
    \caption{Comparing TTFT against different request load patterns of CACE$^{\ast\ast}$, CACE$^{-P4}$, and LRU for RQ2.}
    \label{fig:rq2_ttft}
\end{figure}

\begin{figure}[ht]
    \centering
    \includegraphics[width=\linewidth]{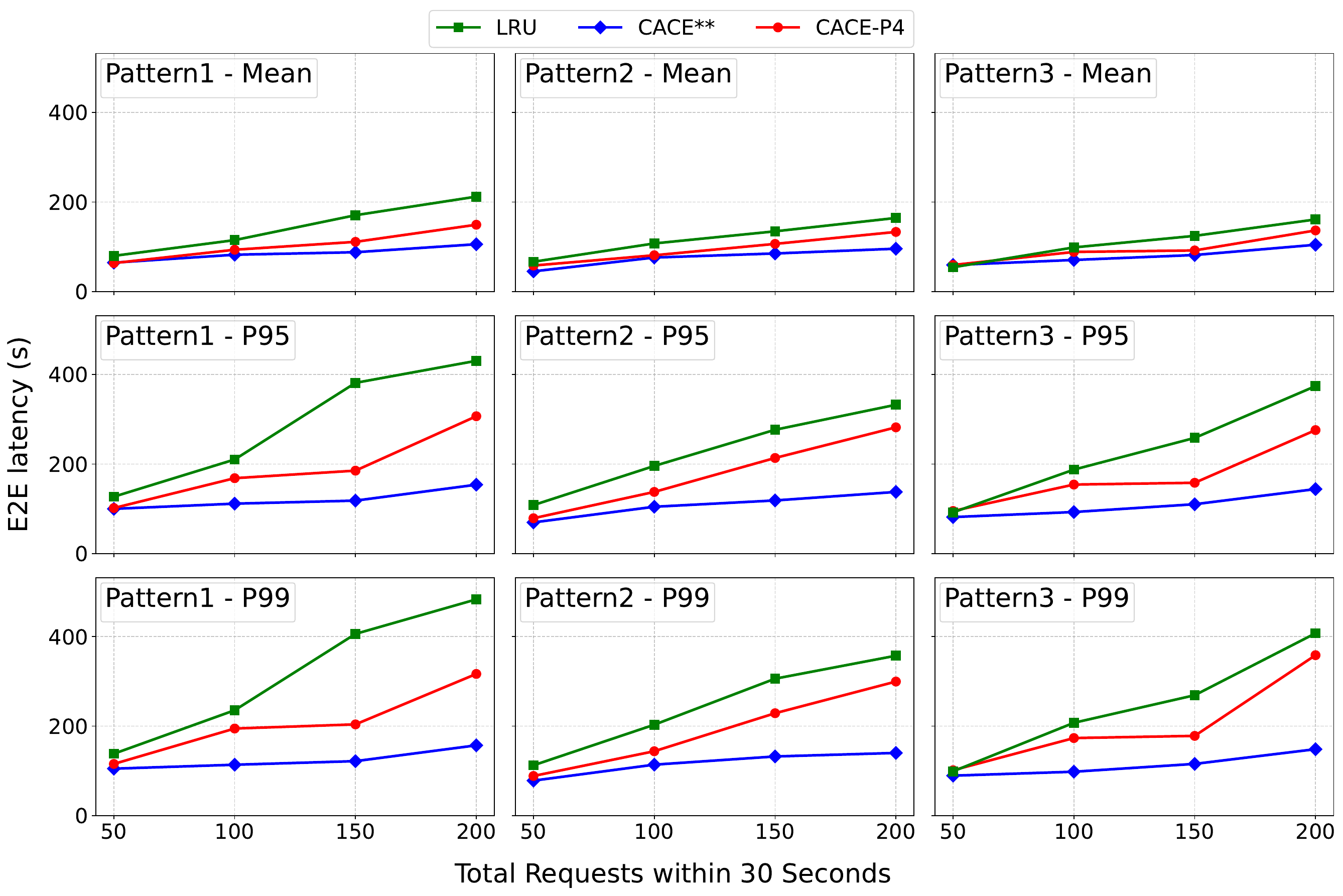}
    \caption{Comparing E2E latency against different request load patterns of CACE$^{\ast\ast}$, CACE$^{-P4}$, and LRU for RQ2.}
    \label{fig:rq2_e2e}
\end{figure}

\begin{figure}[ht]
    \centering
    \includegraphics[width=\linewidth]{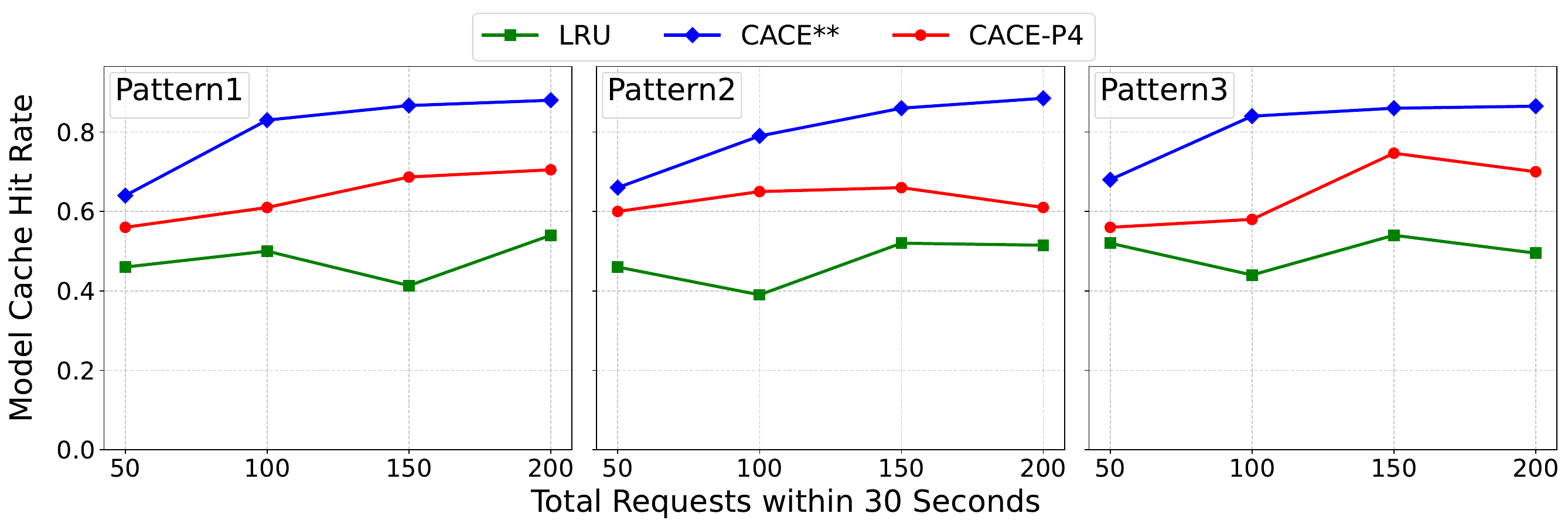}
    \caption{Comparing Model Cache Hit Rate against different request load patterns of CACE$^{\ast\ast}$, CACE$^{-P4}$, and LRU for RQ2.}
    \label{fig:rq2_cache_hit_rate}
\end{figure}

Figures~\ref{fig:rq2_ttft}–\ref{fig:rq2_cache_hit_rate} compare three eviction strategies: baseline LRU, CACE$^{-P4}$ (without Task Criticality), and CACE$^{\ast\ast}$ (all four factors enabled), across three workload patterns. The results confirm that adding the Task Criticality factor improves latency metrics while preserving the eviction benefits observed in RQ1.

Figures~\ref{fig:rq2_ttft}–\ref{fig:rq2_cache_hit_rate} show that enabling Task Criticality (CACE$^{\ast\ast}$) significantly improves latency and cache efficiency across all workload patterns. Compared to LRU, CACE$^{\ast\ast}$ lowers mean TTFT by up to 70\%, and tail latencies by as much as 80\%, with similar improvements in E2E latency, up to 37\% on average and 59\% at the 99th percentile. These latency gains are supported by higher Model Cache Hit Rates, rising to 0.85–0.86 under mixed workloads and reducing model evictions by 55\%. In all cases, CACE$^{-P4}$ (without P4) performs better than LRU, but CACE$^{\ast\ast}$ provides the most consistent and substantial benefits.

\begin{tcolorbox}[colback=white, colframe=black, boxrule=1pt, arc=0pt, left=10pt, right=10pt, top=10pt, bottom=10pt]
\textbf{Findings:} Including task–criticality in the eviction score reduces TTFT by up to 70 \% and E2E latency by up to 37 \%, while boosting Model Cache Hit Rate to 0.86 and halving model load overhead compared with a pure LRU eviction.

\textbf{Implications:} Latency-critical CodeLLM serving systems should prioritize models based on task importance (e.g., expected output length) in addition to recency and reload cost before evicting them from accelerator memory. \textsc{CACE} addresses this by incorporating a Task Criticality factor, which enhances performance for both interactive and long-running tasks. This suggests that multi-factor eviction policies tailored to the latency characteristics of heterogeneous workloads are more effective than single-factor baselines. Future work should investigate how each factor in the eviction score contributes to these gains and whether simpler subsets can produce similar benefits with lower complexity, which motivates the ablation analysis in RQ3.
 
\end{tcolorbox}

\subsection{RQ3: What is the individual impact of each context-
aware factor within the \textsc{CACE} eviction policy?}
\label{subsec:rq3}
While RQ1 and RQ2 evaluate the efficacy of \textsc{CACE}'s eviction scoring strategy under various workload patterns, it is crucial to understand the contribution of each factor within the eviction score formulation. A systematic ablation study allows us to isolate and quantify the contribution of each factor to the eviction score by removing one at a time and evaluating the impact on key performance metrics. 
\subsubsection{Hypothesis}

We hypothesize that disabling individual factors of the eviction score will degrade performance, with certain factors (e.g., Task Criticality) contributing more significantly to metrics like Model Cache Hit Rate and TTFT/E2E latency reduction than others.

We expect:
\begin{itemize}
    \item A decrease in Model Cache Hit Rate when key factors are omitted.
    \item An increase in both TTFT and E2E latency due to suboptimal eviction decisions when key factors are omitted.
    \item Varying degrees of performance degradation depending on the removed factors.
\end{itemize}

\subsubsection{Results}

We reused the same system architecture, datasets, and three workload patterns as described in Section~\ref{subsec:case_study_setup} to ensure consistency. 

As for the request arrival pattern, we use Pattern 3 (Language Popularity Skewed Distribution), as this reflects the real-world language usage with \textbf{IDE Heavy} workload.
In each ablation variant, we remove one factor from the eviction score computation while keeping the other three active. This setup enables a comparative analysis to quantify the individual contribution of each factor toward reducing model evictions and improving model retention in accelerator memory. Since model eviction directly influences the frequency of cold starts, we also examined the downstream effect on latency, specifically, TTFT for code completion tasks and E2E latency for code reasoning tasks. We captured the same evaluation metrics as RQ2 (TTFT, E2E Latency, and Model Cache Hit Rate).

\begin{figure}[h]
    \centering
    \includegraphics[width=\linewidth]{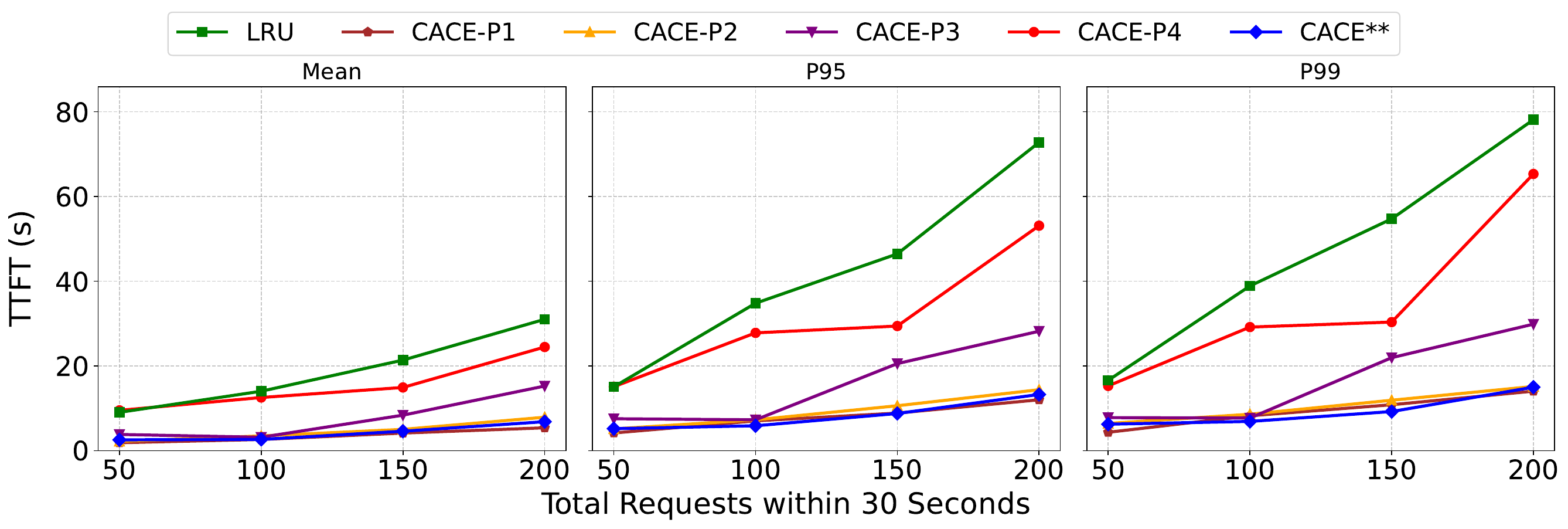}
    \caption{Comparing TTFT of different variants of \textsc{CACE}.}
    \label{fig:rq3_ttft}
\end{figure}

\begin{figure}[h]
    \centering
    \includegraphics[width=\linewidth]{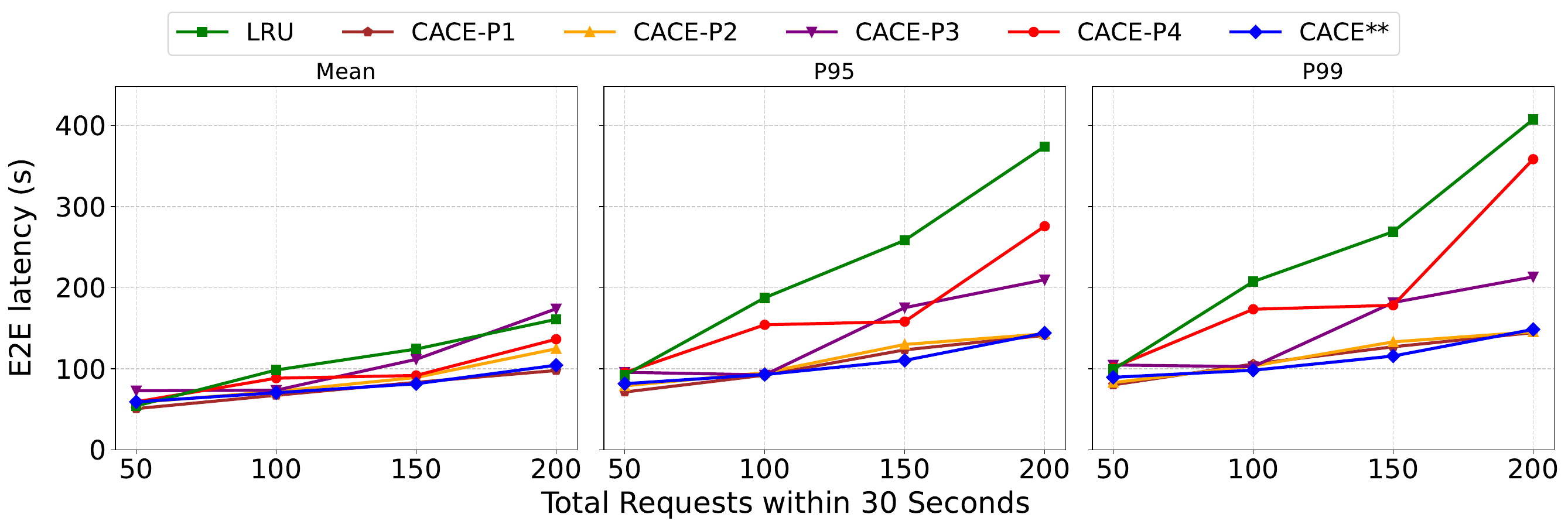}
    \caption{Comparing E2E latency of different variants of \textsc{CACE}.}
    \label{fig:rq3_e2e}
\end{figure}

\begin{figure}[h]
    \centering
    \includegraphics[width=\linewidth]{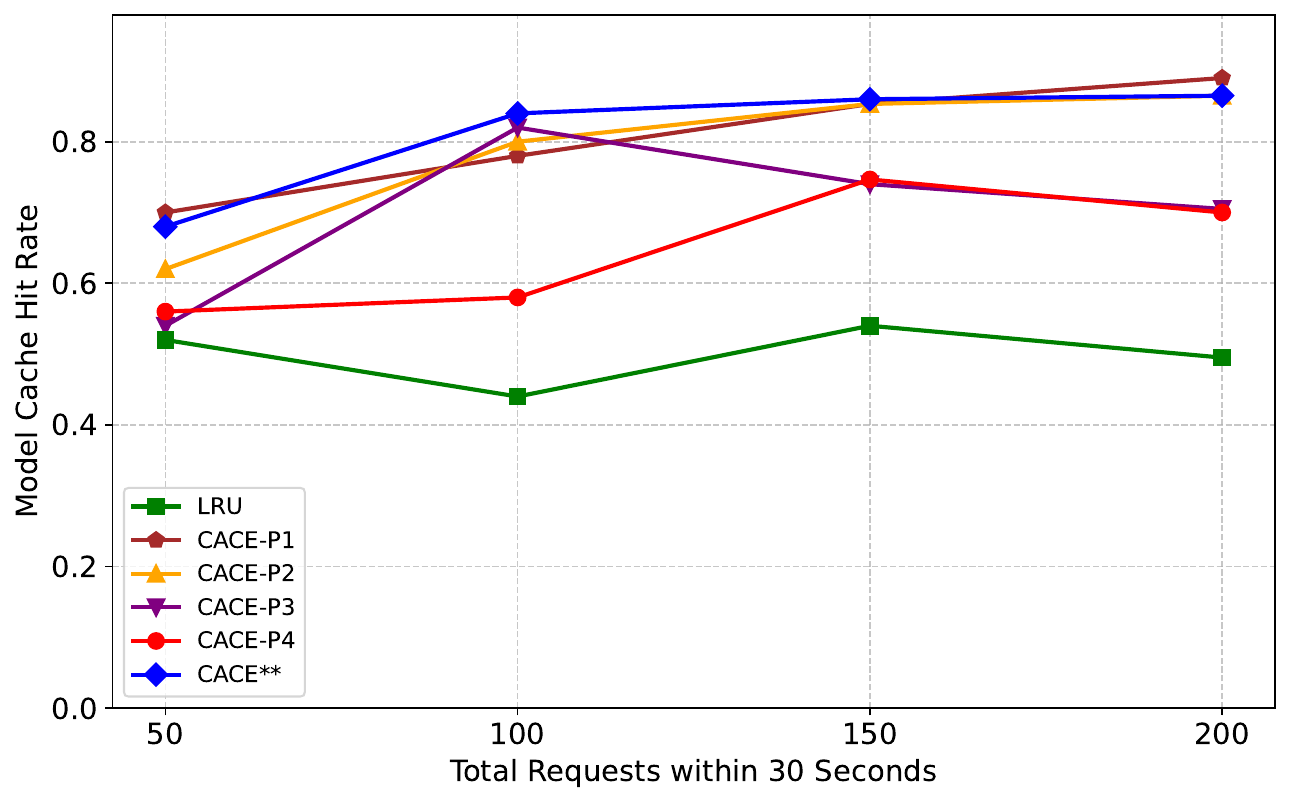}
    \caption{Comparing Model Cache Hit Rate of different variants of \textsc{CACE}.}
    \label{fig:rq3_cache_hit_rate}
\end{figure}

Figures~\ref{fig:rq3_ttft}, \ref{fig:rq3_e2e}, and \ref{fig:rq3_cache_hit_rate} compare CACE$^{\ast\ast}$ with four ablated variants, {CACE$^{-P1}$}, {CACE$^{-P2}$}, {CACE$^{-P3}$}, and {CACE$^{-P4}$}, each obtained by disabling one factor in the eviction score.

Disabling the Task Criticality ({CACE$^{-P4}$}) leads to the largest degradation in Model Cache Hit Rate, dropping from \textbf{0.85} (CACE$^{\ast\ast}$) to \textbf{0.7} (a \textbf{24\%} decrease), and causes the worst latency penalty, with P95 TTFT rising by \textbf{72\%} and P95 E2E latency increasing by \textbf{47}. Removing Future Demand forecasting ({CACE$^{-P3}$}) also substantially harms performance, reducing Model Cache Hit Rate by \textbf{22\%} and increasing P95 TTFT by \textbf{15s}. In contrast, eliminating Recency ({CACE$^{-P1}$}) or Reload Time ({CACE$^{-P2}$}) has only minor effects on both Model Cache Hit Rate and latency. These results underscore that Task Criticality and Future Demand are the dominant factors in effective eviction decisions.

\begin{tcolorbox}[colback=white, colframe=black, boxrule=1pt, arc=0pt, left=10pt, right=10pt, top=10pt, bottom=10pt]
\textbf{Findings:} Ablation results reveal that the Task Criticality factor (P4) is the most influential in improving Model Cache Hit Rate and reducing latency. Disabling it leads to a significant drop in Model Cache Hit Rate and a substantial increase in P95 TTFT, more than any other factor. Future Demand forecasting (P3) is the second most impactful, while recency (P1) and model load time (P2) have comparatively smaller effects.

\textbf{Implications:} The eviction score benefits most from incorporating Task Criticality and Future Demand factors. While recency and reload cost offer modest gains, prioritizing models based on expected output length and future reuse is key to effective model retention in the accelerator memory. Designers can focus on these high-impact factors to optimize performance with manageable complexity. Future work could explore adaptive weighting or dynamic tuning to improve \textsc{CACE}'s responsiveness under changing workloads.

\end{tcolorbox}

\section{Discussion and Future Directions}
\label{sec:discussion}
The experimental insights from our evaluation highlight several promising directions for enhancing the design and implementation of model eviction policies for self-hosted CodeLLM serving systems. Here we outline key areas for future exploration.

\subsection{CPU Memory as a Secondary Cache Layer}

Our current \textsc{CACE} eviction strategy manages models directly between disk storage and accelerator memory. Introducing CPU memory as an intermediate caching layer could significantly reduce reload latencies since model transfers from CPU RAM to accelerator memory are substantially faster than disk-to-accelerator transfers. Models frequently evicted due to accelerator constraints could remain resident in CPU memory, ensuring quicker reload times upon subsequent requests. However, effectively integrating this additional caching layer would introduce new complexities such as accounting for CPU cache capacity, model sizes, and expected reuse frequency. Future work could explore adaptive heuristics or machine learning-based strategies to balance eviction decisions between accelerator, CPU, and disk storage efficiently.

\subsection{\textsc{CACE} for Heterogeneous Accelerator Clusters}

Extending \textsc{CACE} to distributed heterogeneous clusters containing diverse accelerator hardware represents another promising direction. In heterogeneous environments, less urgent or resource-intensive models could temporarily migrate to less performant accelerators or idle nodes, enhancing overall resource utilization. This approach requires sophisticated profiling of inter-node network latency, bandwidth costs, model load time variations, and cluster topology. Future iterations of \textsc{CACE} could incorporate dynamic scheduling strategies informed by real-time network conditions, accelerator capabilities, and task latency sensitivity. Such an approach has the potential to significantly minimize cold-start latencies and improve resource allocation efficiency across heterogeneous distributed deployments.

\subsection{Improved Predictions of Task Criticality through Token Length Estimation}

Accurate prediction of token generation lengths is crucial to the precision of Task Criticality assessments in \textsc{CACE}. Currently, empirical methods (e.g., ExeGPT~\cite{oh2024exegpt}) estimate expected output lengths based on historical model performance data, yet these estimates vary significantly with model architectures and specific coding scenarios. Future work could explore hybrid approaches combining empirical measurements with explicit developer-provided estimates, enhancing prediction accuracy and enabling more proactive eviction decisions. Integrating more precise token-length forecasts or direct user input promises to further improve latency outcomes and optimize resource utilization.

\subsection{Modeling Developer Behavior and Usage Patterns}

Finally, eviction strategies could be further enhanced by explicitly incorporating developer behavior and usage patterns into eviction decisions. Capturing historical developer interaction data, such as typical model usage sequences, working-hour patterns, and task correlations, could enable more accurate forecasts of model demand. By predicting recurring developer interactions, future systems could proactively prefetch necessary models or evict models less likely to be used during certain periods. This direction could substantially reduce cold-start latency, improving the developer experience by better aligning model availability with actual developer workflows.

\section{Related work}
\label{sec:related_work}
Efficient serving of large language models (LLMs), including specialized models for coding tasks (CodeLLMs), has become a central research focus due to their significant computational and memory demands. Prior work can be categorized into three major strategies: full model eviction, KV-cache offloading and paging, and CPU–Accelerator orchestration for Mixture-of-Experts models.

\subsection{Full Model Multiplexing on Shared Accelerators}

Model-multiplexing frameworks load, evict, and route multiple models on a shared accelerator pool to maximize accelerator utilization. Ray Serve~\cite{ray_multiplex} supports fractional-accelerator allocation and concurrent model hosting; its eviction policy is a simple Least-Recently-Used (LRU) heuristic designed for generic ML workloads. Ollama~\cite{ollama} offers an easy-to-deploy, local multiplexing solution, again relying on recency-based eviction and thus incurring cold-start delays as concurrency grows. ModelMesh~\cite{model_mesh}, the engine behind KServe’s “serverless” back-end, targets serving with lazy, on-demand model loading and an LRU-style cache to fit thousands of models into limited accelerator pools. FaaSwap~\cite{yu2024faaswap} improves accelerator sharing in serverless inference by keeping model checkpoints in CPU RAM and evicting them into accelerators on demand; although it reduces eviction latency relative to disk, the decision is still recency-driven and agnostic to task latency requirements or model reload cost. ServerlessLLM~\cite{fu2024serverlessllm} extends this idea with optimized checkpoint storage and faster staging, yet likewise triggers checkpoint loading without considering the heterogeneity of CodeLLM tasks. Snowflake’s model‐hotswapping service~\cite{snowflake} incorporates throughput-based thresholds but similarly overlooks task-level latency sensitivity.

Despite engineering advances, these systems share a core limitation: their eviction logic ignores model load cost and task-specific latency. Evicting a large reasoning model can cost hundreds of milliseconds more than a small completion model, yet pure-LRU or throughput heuristics treat them identically, increasing TTFT or E2E latency under mixed IDE workloads, as shown in the experiments.

\subsection{Single Model Optimization with KV-Cache Offloading and Memory Paging}

A parallel line of work tackles accelerator memory pressure within a single model instance by virtualizing the key–value (KV) cache or paging model weights between device and host memory. PagedAttention in vLLM~\cite{kwon2023vllm} slices the KV cache into fixed-size pages and dynamically evict them between accelerator and CPU RAM, achieving high throughput for long-context inference without increasing model-level eviction frequency. Pie introduces performance-transparent offloading that migrates KV blocks on demand while overlapping transfer and computation to hide latency~\cite{xu2024pie}. FlexGen~\cite{sheng2023flexgen} further explores weight and KV offloading across accelerator, CPU, and even non-volatile memory express (NVMe) tiers to enable very large models on commodity hardware.

These techniques operate below the granularity of whole-model eviction: they optimize memory footprints for a single loaded model and assume the model itself remains resident on the accelerator. Consequently, they do not decide which model to unload when accelerator memory fills, nor do they account for task-specific latency. \textsc{CACE} is complementary: it manages model-level residency across many CodeLLMs, integrating Reloading Cost, Future Demand, and Task Criticality. In practice, a serving stack could combine PagedAttention (for intra-model paging) with \textsc{CACE} (for inter-model eviction), achieving both fine-grained memory efficiency and context-aware model selection.

\subsection{CPU–Accelerator Orchestration for Mixture-of-Experts Models}
\label{sec:related_moe}

Mixture-of-Experts (MoE) architectures partition a large LLM into dozens to thousands of “expert” sub-networks, of which only a few are activated per token \cite{chen2022towards}. This sparse activation reduces floating-point operations per second (FLOPs) but introduces new orchestration challenges: which experts should stay in accelerator memory, and when to migrate cold experts from CPU or NVMe storage.

Fiddler~\cite{kamahori2024fiddler} targets inference for mega-scale MoEs whose total parameters exceed accelerator capacity. Fiddler keeps frequently selected experts resident on accelerators and pre-fetches long-tail experts from CPU RAM based on gating probabilities. DeepSpeed-MoE Inference~\cite{rajbhandari2022deepspeed} similarly employs dynamic expert evictions and kernel-fusion optimizations to serve billion-parameter MoEs on heterogeneous clusters while maintaining sub-second latency.
 
MoE orchestration addresses intra-model weight placement: experts are fragments of the \textit{same} model and share an input pipeline, whereas \textsc{CACE} manages inter-model eviction across many distinct CodeLLMs. While MoE systems optimize which experts reside on the accelerator, they assume a fixed set of models already loaded; they do not decide which entire model family (Python-completion vs. Java-reasoning, for instance) deserves residency when accelerator memory is scarce. Hence \textsc{CACE} is complementary: it determines which full CodeLLMs remain hot, whereas Fiddler-style schedulers can further optimize expert placement within those resident models.

\section{Threats to validity}
\label{sec:threat_to_validity}
In this section, we discuss the threats to validity.

\subsection{Internal Validity}
Our study assumes a single active instance of each CodeLLM in accelerator memory. In practical settings, enterprises often deploy multiple replicas to handle concurrent requests, potentially affecting eviction dynamics. Additionally, our evaluation uses a controlled, single-node setup, simplifying factors like distributed scheduling and real-time metadata accuracy. Any hidden interaction or simplification could influence the observed performance benefits, thus, future evaluations should explore replica-aware eviction policies and multi-node environments.

\subsection{External Validity}
We evaluated two primary task categories: TTFT-sensitive code completion and E2E-sensitive code reasoning. Real-world workloads may include additional tasks such as documentation generation or code refactoring, presenting distinct latency characteristics. Furthermore, our experiments employed a homogeneous hardware setup with identical accelerators and a curated set of CodeLLMs, while actual deployments typically involve diverse hardware and broader model varieties. Thus, further validation across heterogeneous hardware and workloads would strengthen the validity of our findings.

\section{Conclusion}
\label{sec:conclusion}
AI-assisted coding with CodeLLMs has become an essential part of modern software development workflows. To meet privacy, security, and latency requirements, many enterprises are adopting self-hosted solutions for deploying and serving CodeLLMs. However, the growing number and diversity of available CodeLLMs, coupled with the limited accelerator memory, pose significant challenges for efficiently managing model resources, particularly under latency-critical workloads like code completion and code reasoning.

In this paper, we presented a novel model eviction approach that goes beyond traditional LRU policies by incorporating multiple sources of information, including request queue insights, sliding window tracking of recent model usage, future usage demand, model load times, and task criticality based on the expected output length. By integrating these factors, our proposed policy makes more informed eviction decisions that reduce unnecessary data movements and improve response latency for key developer tasks.

We evaluated our approach using realistic developer workloads and showed that it can reduce both model cache misses and overall latency compared to baseline strategies. While our work focuses on single-instance deployments and isolates the eviction logic from full scheduling pipelines, it lays the groundwork for more advanced memory management strategies in production-grade CodeLLM serving systems.

In future work, we plan to extend our approach to multi-instance and distributed settings, heterogeneous accelerators, and utilize CPU memory as a secondary cache layer. We also aim to explore the dynamic adaptation of eviction thresholds based on workload patterns and resource availability.

\section{Disclaimer}
\label{sec:disclaimer}
Any opinions, findings, conclusions, or recommendations expressed in this material are those of the author(s) and do not reflect the views of Huawei. Also, ChatGPT-4.0 was used for copy-editing. All experiments, analysis, writing, and results were performed by the authors, who also thoroughly reviewed the final content. This complies with IEEE and ACM policies on AI use in publications.

\bibliographystyle{IEEEtran}
\bibliography{reference}

\end{document}